\documentstyle[11pt]{article}
\input epsf
\begin{document}

\thispagestyle{empty}
\addtolength{\baselineskip}{0.25\baselineskip}
\rightline{CALT-68-1936}
\rightline{SNUTP-94-67}
\rightline{hep-th/9407173}
\vskip 1.5cm
\centerline{\Large\bf Toward One-Loop Tunneling Rates of}
\centerline{\Large\bf Near-Extremal Magnetic Black Hole Pair-Production}

\vskip 1cm
\centerline{{ Piljin Yi}\footnote{Address
after September 1, 1994: Dept. of
Physics, Columbia University, 538 West 120th St. New York, NY 10027, U.S.A.}
\footnote{e-mail address: piljin@theory.caltech.edu}}
\vskip 3mm
\centerline{452-48 California Institute of Technology}
\centerline{ Pasadena, CA 91125, U.S.A.}
\centerline{and}
\centerline{Center for Theoretical Physics}
\centerline{Seoul National University, Seoul 151-742, Korea}
\vskip 1.5cm
\centerline{ABSTRACT}
\vskip 5mm
\begin{quote}
\noindent
Pair-production of magnetic Reissner-Nordstr\"{o}m black holes (of charges
$\pm q$) was recently studied in the leading WKB approximation. Here, we
consider generic quantum fluctuations in the corresponding instanton geometry
given by the Euclidean Ernst metric, in order to simulate the behaviour of
the one-loop tunneling rate. A detailed study of the Ernst metric suggests
that for sufficiently weak field $B$, the problem can be reduced to that
of quantum fluctuations around a single near-extremal Euclidean black hole
in thermal equilibrium with a heat bath of finite size. After appropriate
renormalization procedures, typical one-loop contributions to the WKB exponent
are shown to be inversely proportional to $B$, as $B\rightarrow 0$, indicating
that the leading Schwinger term is corrected by a small fraction $\sim \hbar
/q^2$. We demonstrate that this correction to the Schwinger term is actually
due to a semiclassical shift of the black hole mass-to-charge ratio that
persists even in the extremal limit. Finally we discuss a few loose ends.
\end{quote}
\newpage

\section{Motivation}

Recently, the pair-production of oppositely charged magnetic black holes
in a background magnetic field, has been studied to the leading WKB
approximation \cite{ERNST}\cite{GGS}\cite{GAUNT}\cite{GG}.
The instanton mediating the tunneling process is found to be the
Euclidean section of the so-called Ernst metric, and the Euclidean
action thereof has been calculated exactly.

One cannot emphasize too much the importance of this process in the context
of black hole quantum physics. A crucial issue in the context of the black
hole information puzzle \cite{Info} is how a potential degeneracy of the black
hole configuration would or would not show up in scattering processes
\cite{remnants}. For instance, it is a matter of some controversy whether it
is possible for remnants with an infinite degeneracy to be pair-produced only
with finite amplitude.

A salient feature of the leading WKB estimate above is that it shows an
enhancement factor of $e^{{\cal S}_{BH}}$ over that of monopole
pair-production,
where ${\cal S}_{BH}$ is the Bekenstein-Hawking entropy \cite{BEK},
seemingly suggesting that the black hole have degeneracy $e^{{\cal S}_{BH}}$
\cite{GGS}.
But this result is more puzzling than clarifying, for one would normally
expect to find the degeneracy, if any, not from the Euclidean action but
at the next order WKB where one explores small deviations from the fixed
instanton configuration. This naturally leads us to next-to-leading order
WKB estimate which we want to explore in this article.

However, our goal here is rather modest. The primary concern here is to
determine if there is anything special about the one-loop pair-production
rate {\it in the extremal limit} which in this context has an alternative
description, namely the weak field limit:
As emphasized in the recent studies and as will be reiterated
in section 2, the temperature of the pair-produced black hole is proportional
to the background magnetic field strength $B$. Hence, we can concentrate
on small $B$ for the purpose of studying near-extremal cases.

First, let us recall the behaviour of the leading exponent \cite{GGS},
simply given by the minus of the Euclidean action divided by $\hbar$
\cite{ColeWKB}.\footnote{In this article,
we use the geometrized unit $c=G=1$, unless specified otherwise.}
\begin{equation}
-\frac{S_{E}}{\hbar}=-\frac{\pi q}{\hbar B}+\cdots \label{eq:Schwinger}
\end{equation}
$B$ is the background magnetic field strength as above, while $q$ is the
absolute value of the magnetic charge of the black hole.
The first term is easily recognized as the Schwinger term that also appears
in the monopole pair-production \cite{MONO},
while the ellipsis denotes terms of higher order
in $qB$, and includes the entropy term $+{\cal S}_{BH}$.

In general, since the one-loop correction to this exponent must be
independent of $\hbar$ by definition, it has to be a dimensionless function
of the parameters of the classical Euclidean instanton, and thus is a
function of $qB=qBG$ only.\footnote{That is, up to the contributions
from the zero-mode sector which we shall be forced to disregard in this first
attempt.} The question is then how strongly it depends on $qB$.  One
interesting property of Reissner-Nordstr\"{o}m black holes is that the
distance to the event horizon diverges along any space-like geodesic
in the extremal limit \cite{Trivedi2},
implying that the physical distance to the Euclidean black hole horizon of the
instanton also diverges in the weak field limit. Furthermore, the periodicity
of an ``asymptotic'' Euclidean time coordinate also diverges owing to the
vanishing black hole temperature. Put together, a possible implication is
that increasing number of quantum fluctuations contribute to the tunneling
process, and that the one-loop correction to the exponent here might have
a very strong $B$-dependence.

For instance, if the additive one-loop correction to the exponent
like $(qB)^{-1-\epsilon}$ with any positive $\epsilon$, the exponent
for near-extremal black hole pair-production would no longer be dominated by
${-S_{E}/\hbar}$, signaling a breakdown of the
semiclassical method. In this article, we wish to investigate exactly how
strongly this one-loop correction scales as $qB\rightarrow 0$.

In the next two sections, we will study the instanton near the horizons,
as a preliminary step. We find through this investigation that, for the
purpose of estimating the leading weak field behaviour of the one-loop
correction, it is sufficient to study quantum fields for a single {\it
truncated} Euclidean black hole in thermal equilibrium with a  heat bath.
This happens because the two lengthscales of the instanton, one associated
with the black hole while the other with the background Melvin universe, are
vastly different from each other for the pair-production of near-extremal
black holes. Furthermore, the size of the {\it truncated} black hole shall be
easily seen to grow indefinitely in the limit of vanishing $B$.
In these two sections, we borrowed heavily, from references
\cite{GAUNT} and \cite{GG}, various results regarding the classical properties
of the instanton.

\vskip 5mm
In order to estimate the genuine one-loop contribution to the tunneling
rate, one must expand the action around the instanton background
and isolate differential operators $Q_a$'s governing the gravitational and the
electromagnetic fluctuations as well as various Fadeev-Popov-like ghost
fields, determinants of which enter the prefactor.
\begin{equation}
\hbox{tunneling rate} \simeq \prod_{a} {\cal N}_a \;
\exp{(-\frac{S_E}{\hbar})}, \qquad \log {\cal N}_a \sim \log {\rm Det} Q_a
\end{equation}
However, since we are interested only in their qualitative behaviour, it
is reasonable to consider generic second-order operators instead, and for the
rest of the article, we shall analyze one-loop contributions from
various {\it matter} degrees of freedom. (Also, we will consider chargeless
fluctuations only in this article.) Note that by this substitution we
completely destroy the information on the zero-mode sector as well as
the negative-mode of the gravitational and electromagnetic fluctuations.
Therefore, we are actually considering the contribution from the positive mode
sectors only.

\vskip 5mm
In section 4, we take on the first examples: those of generic massive quantum
fluctuations. For this case, we estimate the determinant of a general elliptic
operator with large positive mass term by performing the Schwinger-DeWitt
expansion \cite{SD}. After a renormalization procedure, we find the one-loop
correction to the exponent diverge like $\sim 1/qB$, as $qB\rightarrow
0$. Note that this has the same $B$ dependence as the Schwinger term shown
above.

To determine whether a similar effect arises in massless cases, we study two
special kinds of massless fluctuations: chargeless Callan-Rubakov modes
\cite{Callan}\cite{Alford}\cite{Holzhey}\cite{theta} and
four-dimensional conformal fluctuations. Because of the non-local nature of
the resulting effective actions, here we need to take into account
the heat bath mentioned above. It is shown in section 5 that chargeless
Callan-Rubakov modes, just as the massive fluctuations, contribute to the
exponent a term $\sim 1/qB$. Furthermore, we estimate the proportionality
coefficient exactly. Calculations for the four-dimensional conformal
fluctuations are a bit more involved, and we employ an alternate method
that is suitable for finding the leading $qB$ dependence.
Performing a dimensional analysis of a metric variation of the effective
action expressed in terms of the one-loop energy-momentum expectation, we
argue that exactly the same $1/qB$ behaviour occurs here as well.

In our conclusion, we demonstrate that this correction
actually originates from a semiclassical shift of the near-extremal
black hole masses. The correspondingly shifted mass-to-charge ratio of the
semiclassically corrected extremal Reissner-Nordstr\"{o}m black holes
was recently studied and estimated in reference \cite{Jaemo}, and is due
to certain residual quantum radiation that exists despite the vanishing
Hawking temperature. The net result is then, we find no large and novel
quantum correction associated with near-extremal black hole pair-production
as far as contributions from the matter sector are concerned. Also we
discuss a few loose ends that need further study.

\section{The Geometry of the Euclidean Ernst Metric in the Weak Field Limit}

Let us first write down the instanton solution of the Einstein-Maxwell
theory \cite{ERNST}, various parameters of which are to be identified
with those of the pair-produced black holes. We follow the coordinate
convention of reference \cite{GAUNT}.
\vskip 1mm
\begin{eqnarray}
g^{(4)}&=&\frac{\Lambda^2}{(Ay-Ax)^2}\bigl(-G(y)dT^2-\frac{dy^2}{G(y)}\bigr)
       +\frac{1}{(Ay-Ax)^2}\bigl(\frac{\Lambda^2}{G(x)}dx^2
                +\frac{G(x)}{\Lambda^2}d\phi^2\bigr) \label{eq:E-E} \\
G(\xi)&=&(1+r_{-}A\xi)(1-\xi^2-r_{+}A\xi^3) \\
\Lambda&=&\Lambda(x,y)=(1+\frac{1}{2} qBx)^2+\frac{B^2}{4(Ay-Ax)^2}G(x)
\end{eqnarray}
\vskip 5mm
\noindent
The solution comes with two Killing coordinates $T$ and $\phi$, where the
latter generates an axial symmetry. The Minkowskian version of this metric
has an interpretation as two magnetic black holes of opposite charges,
accelerating away from each other in the background Melvin universe.
In this section, we are interested in the geometry of this instanton solution
in the weak field limit of small $B$. As will be shown later, the only two
independent parameters are $q\equiv \sqrt{r_+ r_-}$ and
$B$; in the limit $qB\rightarrow 0$,
$q$, $A$, and $B$ are respectively the absolute value of charges, the
acceleration of the black holes, and the magnetic field strength on the
symmetry axis.

\vskip 5mm

Calling the four roots of $G(\xi)$, $\xi_{1}$, $\xi_{2}$, $\xi_{3}$,
$\xi_{4}$ in the ascending order, $y=\xi_{1}$, $\xi_{2}$, $\xi_{3}$
correspond to the inner and the outer horizons and the acceleration
horizon respectively. But along the Euclidean section of real $T$, the
inner horizon at $y=\xi_1$ is not part of the instanton, and the
$y$-coordinate
is restricted to $\xi_{2} \leq y \leq \xi_{3}$, that is between the outer
horizon and the acceleration horizon. Similarly,  $x$, an angular
coordinate, is restricted to $\xi_{3} \leq x \leq \xi_{4}$.

To study the limit $qB \rightarrow 0$, it is convenient to expand $\xi_i$
in terms of $r_{\pm}A$ which are also necessarily small in the same limit,
\begin{eqnarray}
\xi_{1}&=&-\frac{1}{r_- A} \\
\xi_{2}&=&-\frac{1}{r_{+}A}+r_{+}A+ \cdots \nonumber\\
\xi_{3}&=&-1-\frac{r_{+}A}{2} +\cdots \nonumber \\
\xi_{4}&=&+1-\frac{r_{+}A}{2} +\cdots. \nonumber
\end{eqnarray}
There are four potential  conical singularities at the boundaries
of the coordinates $x$ and $y$. Naturally, two of them, say at $x=\xi_{4}$ and
$y=\xi_{3}$, can be removed by adjusting the periods of $\phi$ and $T$, while
the removal of the rest forces certain relationships between the
parameters of the instanton. For instance, the geometry is smooth at
$x=\xi_{3}$ only if $A$ satisfies the following identity \cite{GAUNT},
\begin{equation}
\frac{(r_{+}+r_{-})}{2}A=qB+\cdots  \label{eq:newton}
\end{equation}
which is nothing but the Newton's law, once we identify $(r_+ +r_-)/2$ as the
mass of the black hole.

The final conical singularity at the black hole event horizon $y=\xi_{2}$
is resolved by adjusting $r_{+}-r_-$ according to the following constraint
\cite{ERNST}\cite{GAUNT}.
\begin{equation}
\frac{(r_{+}-r_{-})}{4\pi r_{+}^2}=\frac{A}{2\pi}+\cdots \label{eq:R-H}
\end{equation}
It is easy to recognize that, up to a factor of $\hbar$ on each
side, the right-hand-side is the Unruh temperature \cite{Unruh}\cite{Wald}
of an observer with
acceleration $A$, and the left is the Hawking temperature
\cite{Hawking} $T_{BH}$ of a
Reissner-Nordstr\"{o}m black hole with horizon radii $r_{\pm}$.

\vskip 5mm
This suggests that the pair-produced objects are near-extremal magnetic
Reissner-Nordstr\"{o}m black holes of temperature $\simeq B\hbar/2\pi$,
accelerating
away from each other. With the present form of the metric that covers the
whole Euclidean space, however, it is difficult to see whether the black holes
are indeed of Reissner-Nordstr\"{o}m type. For instance, the solution
(\ref{eq:E-E}) lacks the spherical symmetry that is characteristic of
such black holes, owing to the background magnetic field that forces the
acceleration. But since the background magnetic field and the black holes are
described by two separate lengthscales  $B^{-1}$ and $q$, the effect
of the background magnetic field on the geometry near black holes may be
ignored in the weak field limit $B^{-1} \gg q$, and a suitable expansion
in terms of $qB$ should reveal the hidden black hole geometry.

Near the Euclidean black hole whose horizon is at $|y|= -\xi_{3}\simeq
1/r_{+}A \simeq 1/qB \gg 1$, then, we might as well expand the metric
in terms of $1/|y|$. It is particularly convenient to perform the following
coordinate transformations \cite{GAUNT}.
\begin{equation}
\tau=T/A,\qquad r=-\frac{1}{Ay} \label{eq:NEW}
\end{equation}
Rewriting the functions in the metric coefficients, in terms of $r$ and $x$,
\begin{eqnarray}
(Ay-Ax)^2&=&\frac{1}{r^2}(1 +rAx)^2, \nonumber \\
G(y)&=&(1-\frac{r_{-}}{r})(r^2A^2-1+\frac{r_{+}}{r})
\frac{1}{r^2A^2},\nonumber \\
G(x)&=&(1+r_{-}Ax)(1-x^2-r_{+}Ax^3), \nonumber \\
\Lambda(x,y)&=&(1+\frac{qB}{2}x)^2+ \frac{r^2B^2}{4(1+rAx)^2}G(x), \nonumber
\end{eqnarray}
\noindent
and retaining the leading nonvanishing term in each expression as $r_\pm A
\rightarrow 0$,
\begin{eqnarray}
(Ay-Ax)^2&=&\frac{1}{r^2}+\cdots,\\
G(y)&=&-\frac{1}{r^2A^2}(1-\frac{r_{-}}{r})(1-\frac{r_{+}}{r})+\cdots,  \\
G(x)&=&1-x^2+\cdots, \\
\Lambda&=&1+\cdots.
\end{eqnarray}
Recall that the angular coordinate $x$ is confined to $[\xi_{3},\xi_{4}]
\simeq [-1,1]$.

Finally inserting these approximate expressions back to the metric
(\ref{eq:E-E}), and introducing an angular coordinate $\cos\theta=x$,
we recover the Reissner-Nordstr\"{o}m geometry to the leading nonvanishing
order, as expected.
\begin{eqnarray}
g^{(4)}&=&r^2\,\bigl( \frac{1}{r^2A^2}(1-\frac{r_{-}}{r})(1-\frac{r_{+}}{r})
         \,A^2 d\tau^2+\frac{r^2A^2}{(1-\frac{r_{-}}{r})(1-\frac{r_{+}}{r})}
         \,d(\frac{-1}{Ar})^2 \bigl) \nonumber \\
       &+&r^2\,\bigl(\frac{dx^2}{1-x^2}+(1-x^2)\,d\phi^2\bigr)
         +\cdots \nonumber \\
       &=&(1-\frac{r_{-}}{r})(1-\frac{r_{+}}{r})\,d\tau^2+
         \frac{1}{(1-\frac{r_{-}}{r})(1-\frac{r_{+}}{r})}\,dr^2+
         r^2\,(d\theta^2+\sin^2\theta\,  d\phi^2)+\cdots. \label{eq:R-N}
\end{eqnarray}
Note that this form of the metric provides a good approximation to the
instanton geometry only when the radius $r$ is much smaller than the scale
set by the magnetic field strength, namely when $r \ll B^{-1}$.

In any case, the discussions above clearly show that, in the weak field
limit $qB\rightarrow 0$, this instanton mediates a quantum tunneling process
that pair-produces magnetic Reissner-Nordstr\"{o}m black holes of Hawking
temperature $\simeq \hbar B/2\pi$. Since there is only one
Euclidean black hole horizon of the instanton, these two black holes must be
identified along their bifurcation surfaces, thus forming a single Euclidean
black hole when continued to Euclidean space.

\section{One-Loop Correction, Vacua and the Accelerating Black Holes}

In the previous section we emphasized that, in the weak field limit, the
instanton geometry near the Euclidean black hole horizon is that of a
{\it single} near-extremal Euclidean black hole. In the figure 1, the
region described by the approximate metric (\ref{eq:R-N}) corresponds to the
bottom ``cup'' attached to the background Melvin space as depicted by the
top ``sheet.'' Clearly the Euclidean black hole region is not part of the
background Euclidean Melvin universe and  uniquely associated with the
pair-production process.

\vskip 20mm
\begin{center}
\leavevmode
\epsfysize=1.5in \epsfbox{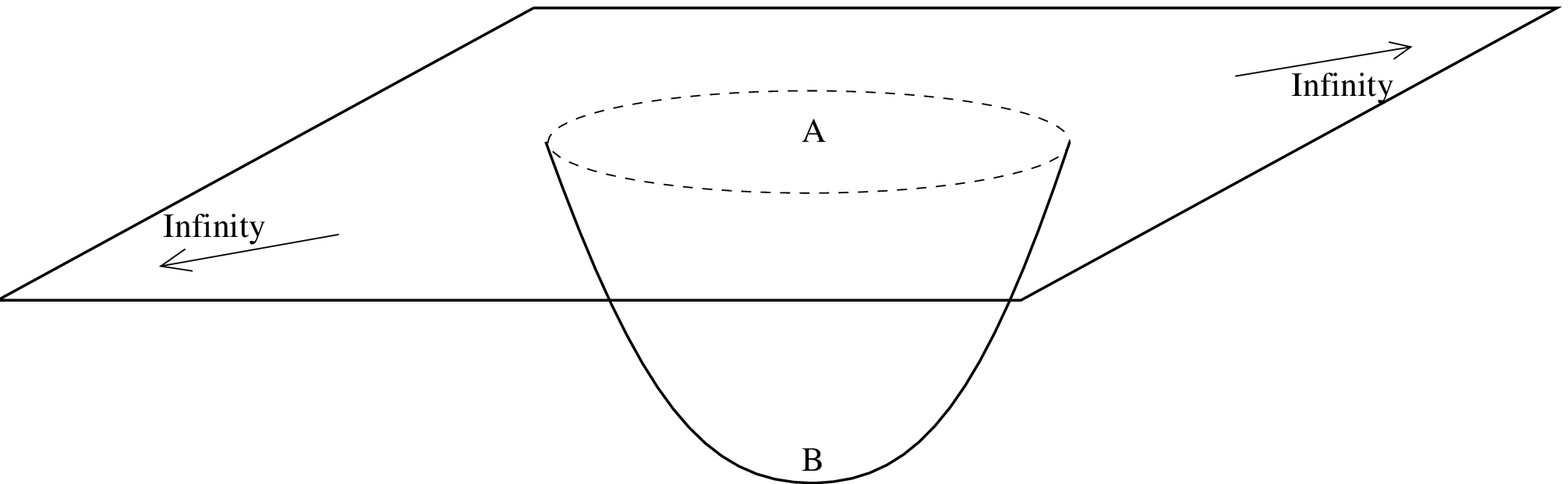}
\end{center}
\vskip 5mm
\begin{quote}
{\bf Figure 1:} {\small A schematic diagram for the Euclidean instantons.
In the weak field limit, the bottom ``cup'' is described by a near-extremal
Euclidean black hole, while the top ``sheet'' is the Euclidean Melvin space.
The acceleration horizon and the black hole horizon are located at points
A and B, respectively. The transitional ``mouth'' region is denoted by the
broken curve where the area of the transverse two-sphere is $\sim 4\pi\,
q^{2/3} B^{-4/3}$. }
\end{quote}
\vskip 10mm

Since the tunneling rate can be written as a ratio of two partition functions,
one associated with the instanton mediating the tunneling process and
the other associated with the background Melvin space, evaluating the
contribution from this black hole region should give us a qualitative
behaviour of the one-loop correction. We are assuming that there
are no large correlations between the bottom ``cup'' and the top ``sheet.''

In principle, therefore, one may perform the necessary mode sum of quantum
fluctuations with supports in the bottom ``cup'' only, in order to
extract the one-loop contribution to the exponent, or alternatively, one
may start with the {\it off-shell} effective action for an arbitrary
background geometry and try to evaluate it on the given geometry,
as we will do throughout this article.
However, if the {\it off-shell} effective action to be evaluated
is a non-local functional of the metric, the on-shell
value of the effective action is a functional not only of the curvatures
and the connections but also of various Green's functions, and it is most
important to perform the calculations in the right vacuum. And this is
where the rest of the instanton is needed.

As illustrated in the figure 1, the instanton geometry
possesses two horizons: the black hole horizon B on the bottom  ``cup''
and the acceleration horizon A on the top ``sheet.''  The surface gravities
of these two horizons are identified to ensure the absence of conical
singularities \cite{ERNST}\cite{GAUNT}, which, according to (\ref{eq:R-H}),
implies that the Hawking radiation of the near-extremal black holes is
balanced against the Rindler heat bath \cite{GG}.

This picture of black holes immersed in Rindler heat baths can also be seen
from the fact that the time-coordinate $T=A\tau$ plays the role of Euclidean
Rindler time. Far away from the black hole horizon $r_+ A\,|y|\ll 1$, the
Ernst metric (\ref{eq:E-E}) can be transformed into the following
Rindler-like form \cite{GG},
\begin{equation}
g^{(4)}\simeq \Lambda^2\,(  \zeta^2\,dT^2+d\zeta^2+d\rho^2)+\Lambda^{-2}\,
\rho^2\,d\phi^2,\qquad \Lambda\simeq 1+\frac{B^2\rho^2}{4}, \label{eq:Rindler}
\end{equation}
by performing the  coordinate transformations
\begin{equation}
\zeta^2= \frac{y^2-1}{A^2(x-y)^2},\qquad \rho^2=\frac{1-x^2}{A^2(x-y)^2}.
\end{equation}
Clearly observers at fixed $\zeta$, $\phi$, and $\rho$ experience
acceleration of $1/\Lambda\zeta$.

On the other hand, for sufficiently small $qB\simeq r_+ A$, there is a
transitional ``mouth'' region $r_+ A\,|y|\ll 1 \ll |y|$,
where the Ernst metric is
fairly well approximated by either of the metrics (\ref{eq:R-N}) and
(\ref{eq:Rindler}). In terms of the coordinates of (\ref{eq:R-N}) and
of (\ref{eq:Rindler}), this transitional region is located at
\begin{equation}
r_+ \ll r \ll \frac{1}{A},\quad \hbox{or} \quad \rho \ll \frac{1}{A}
\;\hbox{and}\; \zeta\simeq \frac{1}{A}. \label{eq:MOUTH}
\end{equation}
Therefore, the ``asymptotic'' observers at large fixed $r$, such that
$r_+\ll r \ll A^{-1}$, are in
fact Rindler observers with the acceleration given by $\zeta^{-1}\simeq A$,
and, to the first nonvanishing order in $qB$, they must find the Hawking
radiation from the black hole in a perfect equilibrium \cite{GG}
with the Rindler heat bath \cite{Unruh}.\footnote{
The fact that the black holes must be in a thermal equilibrium, is certainly
consistent with the interpretation of the instanton as a mediator of the
tunneling process. A smooth transition from the Euclidean instanton to the
Minkowskian Ernst metric, describing pictorially the process of
pair-production,
is possible because the configuration is at rest at the moment
of transition. If we want to extend this picture to the one-loop level,
the configuration at the transition must achieve some sort of semiclassical
equilibrium. By letting the black holes  be in thermal equilibrium with
the Rindler heat-baths, we ensure that
the instanton solution is reliable even to the one-loop level.}

To understand another aspect of this vacuum,  recall that the local
temperature, measured by a ``static'' observer propagating along $\partial/
\partial_T$, is given by the inverse of the proper period of his Euclidean
orbit. A ``static'' observer at large $r\ll A^{-1}$, for example,
has the period $\simeq \hbar/T_{BH}$, as expected from an approximate black
hole geometry (\ref{eq:R-N}), and as emphasized above.
As we climb out of the bottom ``cup'' region, however, the geometry is no
longer that of a Euclidean black hole, but is described by an approximate
Melvin space (\ref{eq:Rindler}).

The obvious result is that the local temperature of this vacuum state vanishes
rapidly ($1/2\pi \zeta \Lambda \rightarrow 0$) as $\zeta\rightarrow \infty$,
owing to extra gravitational red-shifts, similar to those in the Rindler
space. This is in stark contrast with the usual Hartle-Hawking
vacuum around a black hole, where the local temperature approaches the
Hawking temperature $T_{BH}$ asymptotically.\footnote{If the geometry were
that of a Euclidean black hole everywhere, the proper period would approach
the finite value $\hbar/T_{BH}$ as $r\rightarrow \infty$, for the
geometry along the $(\tau,r)$ plane resembles a cylinder asymptotically.}
The upshot from these observations is that this naturally motivated vacuum
looks like the Hartle-Hawking vacuum inside the bottom ``cup,'' yet, outside,
behaves as if it were an ordinary Minkowski vacuum as seen by the family of
accelerating Rindler observers, up to the
modifications due to the background magnetic fields. This tells us that,
among other things, the heat bath associated with the instanton comes with
a natural infrared cut-off given by the size of the bottom ``cup.''

Of course, for the purpose of evaluating the one-loop exponent, we would
have had introduced a similar cut-off in any case, since, as emphasized above,
it is only the bottom ``cup'' region of the instanton (figure 1) that is
associated with the tunneling process and thus contributes to the tunneling
rate. Now what is the size of the bottom ``cup''? Note that, according to
(\ref{eq:MOUTH}) and (\ref{eq:Rindler}), the typical curvature of the
``mouth'' region is roughly $\sim B^2$ due to the nontrivial factor
$\Lambda\simeq 1+B^2\rho^2 /4$. On the other hand, the same region is
described by the black hole metric (\ref{eq:R-N}) at some large value
of $r$, say $r_B$, and the corresponding curvature scale is $\sim
q/r_B^3$. Equating these two scales, we find the approximate value of
the $r$ coordinate along the ``mouth'' region: $r_B \sim (qB^{-2})^{1/3}$.
As a result, whenever $q \ll r_B \ll B^{-1}$, the bottom ``cup'' is
well approximated by a {\it truncated} Euclidean black hole (\ref{eq:R-N})
with $r$ restricted to be smaller than $r_B$, the effective {\it size} of the
heat bath. We shall see in the following sections that the precise value of
$r_B$ is immaterial as far as the leading weak field behaviour is
concerned.

To summarize, we argued that a rough estimate of the one-loop prefactor
${\cal N}_a$ may be obtained by evaluating the corresponding
effective action on a single near-extremal Euclidean black hole
truncated at $r=r_B\sim (qB^{-2})^{1/3}$ and in thermal equilibrium with a
heat bath.

\section{The Leading One-Loop Contribution in the Weak Field Limit:
Massive Fluctuations}

Let us first consider the case where the choice of the vacuum is already
built in. If the fluctuation is massive enough, the calculation of the
effective action can be done in a systematic local expansion in terms
of both curvatures and momentum, known as the Schwinger-DeWitt expansion
\cite{SD}. Since each term of this expansion is a local expression of
the curvatures and the connections, no ambiguity regarding the boundary
condition may arise. Without loss of generality,
take a (bosonic) chargeless fluctuation of mass $\cal M$ and the kinetic
operator $Q+{\cal M}^2$. Then the prefactor ${\cal N}_{\rm massive}$ is
related to the determinant in the following way,
\begin{equation}
{\cal N}_{\rm massive}= {\rm Det}^{-1/2}\,\{Q+{\cal M}^2 \}
\equiv e^{-W},
\end{equation}
where we evaluate the expressions on the right hand side on the given
background geometry. The effective action\footnote{We shall use the same
notation $W$ for various effective actions associated with various quantum
fluctuations considered in this article and also for the {\it  total}
effective action.}  $W$ is most
conveniently expressed in terms of the heat kernel \cite{Heat}.
\begin{equation}
W\equiv \frac{1}{2}{\rm Tr}\log (Q+{\cal M}^2)=-\frac{1}{2}
\int_{\epsilon}^{\infty} \frac{ds}{s}\:{\rm Tr}\,e^{-sQ-s{\cal M}^2}
\end{equation}
The lower limit $\epsilon$ is an ultraviolet cut-off that has the dimension of
length squared. Performing the Schwinger-DeWitt expansion, we find
\begin{equation}
{\rm Tr}\,e^{-sQ-s{\cal M}^2}= \frac{e^{-s{\cal M}^2}}{16\pi^2 s^2}
\sum^{\infty}_{n=0} s^n\, \biggl\{\int dx^4 \sqrt{g}\: {\cal L}^{(n)}\biggr\}.
\label{eq:SdW} \end{equation}
The integral is over the truncated Euclidean black hole of the previous
section. ${\cal L}^{(n)}$ is a polynomial of the curvature tensor $R^\alpha_{
\beta\gamma\delta}$ and its derivatives, and has the same scaling dimension as
$R^n$. In particular, ${\cal L}^{(0)}$ is always identical to $1$,
and ${\cal L}^{(1)}$ is proportional to the curvature scalar.

Obviously the first two terms $n=0$ and $n=1$ contribute to the
renormalization of the cosmological constant and the gravitational constant
respectively, while a logarithmically divergent part of the third ($n=2$) can
be absorbed in the renormalization of  dimension-four operators.

Then, after the usual renormalization process, the remaining finite
contributions from $n\ge 2$ can be integrated over both $s$ and $x$.
Note that, since the integrands are completely independent of $\tau$,
the integral over the Euclidean time simply produces a universal factor
of $\hbar/T_{BH}$ which is the periodicity of $\tau$. On the other hand,
the integrals over the radial and angular coordinates as well as over $s$
induces a power series in $1/q{\cal M}$.\footnote{Since the integrands
vanish very rapidly, the upper limit of the $r$ integration does not
make much difference.}
\begin{equation}
W = \frac{\hbar}{qT_{BH}}\:\biggl\{\sum^{\infty}_{m=0}
\frac{a_m}{({q\cal M})^{2m}}\biggr\}
+\cdots
\end{equation}
Coefficients $a_n$ are constants and the ellipsis denotes terms of
higher order in $qB$. Now using the fact that $B/2\pi \simeq T_{BH}/\hbar$,
\begin{equation}
W=\frac{2\pi}{qB}\:\biggl\{\sum^{\infty}_{m=0} \frac{a_m}{(q{\cal M})^{2m}}
\biggr\}+\cdots \label{eq:SUM}
\end{equation}
Finally denoting by $\sigma/2$ the sum inside the curly bracket, we find
the following exponent of one-loop tunneling rate.
\begin{equation}
-\frac{S_{E}}{\hbar}-W= -\frac{\pi q}{\hbar B}-\frac{2\pi}{qB}\, \biggl\{ \sum
\frac{a_m}{(q{\cal M})^{2m}}\biggr\}+\cdots =-\frac{\pi q}{\hbar B}\biggl(
1+\sigma \frac{\hbar}{ q^2} \biggr)+\cdots. \label{eq:Massive}
\end{equation}
As a consequence, this one-loop effect does not go away in the weak field
limit $B\rightarrow 0$ and represents a multiplicative correction of
the leading Schwinger term by a fraction $\sim \hbar/q^2$, independent
of how small $B$ is.

In a sense, this result is what one would expect from the most naive
dimensional analysis. Restoring the gravitational constant $G$, the fractional
correction above can be rewritten in terms of Planck
length $L_{Planck}$ and the size of the black holes $L_{BH}$.
\begin{equation}
\frac{\hbar W}{S_E}= \sigma\frac{\hbar}{ q^2}+\cdots\;\sim\;
\frac{\hbar G}{q^2 G}\:\simeq\;\biggl( \frac{L_{Planck}}{L_{BH}}\biggr)^2,
\end{equation}
which is the most natural small parameter characterizing the strength of
the quantum fluctuations. But a subtlety arises in our case because there
is another small parameter $qB$. {\it A priori},
the leading one-loop correction can take a more general form.
\begin{equation}
\frac{\hbar W}{S_E}\sim\; \biggl(\frac{1}{qB}\biggr)^k \:\frac{\hbar}{q^2}
\end{equation}
The question is then whether the result (\ref{eq:Massive}) indicating $k=0$
will generalize to other cases. As emphasized in the first section,
the bottom ``cup'' region of figure 1 comes with ever-increasing volume
as $qB\rightarrow 0$, and it is unclear whether there exists an unknown
infrared effect capable of producing such extra factors of $1/qB$. Hence
it is imperative to investigate how massless fluctuations contribute.
\vskip 5mm

One crucial property of the expanded effective action above is that
the integrands ${\cal L}^{(n)}$ are all regular at the Euclidean black hole
horizon. This is in turn guaranteed by the term-by-term locality of the
effective action. For such integrands, the integrations over the radial
and the angular coordinates should be completely determined by the {\it
classical} lengthscales of the black hole, and cannot produce extra factors
of the inverse temperature $\sim 1/qB$; hence the result (\ref{eq:Massive}).

However, the Schwinger-DeWitt expansion is no longer adequate for massless
fluctuations, ${\cal M}=0$, since each term with $n>1$ will cause an
infrared divergence upon the $s$ integration. The formula (\ref{eq:SUM})
shows this pathology manifestly, for each
term of the series inside the curly bracket diverges when $\cal M$ vanishes.
To deal with such massless fluctuations, we need to perform a resummation
over the momentum part of the Schwinger-DeWitt expansion, in favor of a
manifestly non-local expression of the effective action \cite{Covariant}.

In the following two sections, we want to explore the one-loop contributions
from massless fluctuations with proper care of the boundary condition.
But since the
necessary calculations for arbitrary massless fluctuations are hopelessly
difficult, we will specialize to two special cases: massless S-waves
that are effectively 2-D conformal fields, and then, 4-D conformal fields.

\section{The Leading One-Loop Contributions in the Weak Field Limit: Massless
S-Wave Fluctuations}

For most quantum fluctuations around the instanton solution, there
exist potential barriers near the black hole event horizon.
For instance, angular momentum $l$ modes of a minimally coupled massless
scalar find the following potential barrier $V_l$ in the tortoise coordinate
$z$.
\begin{equation}
g=F(z)\,(-dt^2+dz^2)+R^2(z)\,d\Omega^2\qquad \Rightarrow \qquad V_l(z)=
\frac{\partial_z^2 R}{R}+\frac{l(l+1)\, F}{R^2}
\end{equation}
Such potential barriers are especially inhibitive for low energy
excitations which are responsible for the infrared behaviour of the
effective action.

One exception to this is the celebrated Callan-Rubakov modes
\cite{Callan}\cite{Holzhey} in spherically
symmetric magnetic field backgrounds, chargeless combinations of which
propagate effectively as 2-D conformal fields
\cite{Alford}\cite{theta}. While the instanton geometry
is not spherically symmetric everywhere, we have seen that, in the weak field
limit $qB\rightarrow 0$, the spherical symmetry is restored near the
Euclidean black hole horizon, giving us some hope that the contribution from
these uninhibited modes near the Euclidean black hole may capture the
essential physics of the next-to-leading WKB.

For 2-D conformal fields, the corresponding effective actions are exactly known
in explicitly non-local form. A resummation over the momentum part of the
Schwinger-DeWitt expansion turns out to produce a single nonlocal term
\cite{Covariant} known as the Polyakov-Liouville action \cite{Polyakov},
up to a local topological contribution. Accordingly, the
prefactor ${\cal N}_{\rm CR}$ from $N$ such S-wave modes can be written
as following,
\begin{equation}
{\cal N}_{\rm CR}=
e^{-W}=e^{-N S_{PL}},\qquad S_{PL}=\frac{1}{96\pi}\int dx^2\sqrt{g^{(2)}}\,
R^{(2)}\frac{1}{\nabla^2}R^{(2)}\: +a\int dx^2 \sqrt{g^{(2)}}\,R^{(2)}
\end{equation}

\noindent
We included the topological term whose coefficient $a$
depends on the renormalization scheme in two-dimensions.

The form of Polyakov-Liouville action
becomes particularly convenient to handle in conformal  coordinates, since the
scalar curvature $R^{(2)}$ above can be expressed simply by the Laplacian
of the conformal mode. Here, let us choose an asymptotically flat
conformal coordinate.
\begin{equation}
g^{(2)}=F\,(d\tau^2+dz^2) \quad \rightarrow \quad R^{(2)}=-\nabla^2 \log F.
\end{equation}

\noindent
For the case of the Euclidean Reissner-Nordstr\"{o}m black hole, in particular,
$F$ and $z$ are given by
\begin{equation}
F=F(r)\equiv (1-\frac{r_{-}}{r})(1-\frac{r_{+}}{r}), \qquad
z=\int\frac{dr}{F(r)}.
\end{equation}

\noindent
The importance of choosing the conformal gauge can be seen from the following
general relationship, where $h$ is any harmonic function on the given space,
to be fixed by the boundary condition chosen:
\begin{equation}
\frac{1}{\nabla^2}R^{(2)}=-\frac{1}{\nabla^2}\nabla^2 \log F=-\log F+h .
\label{eq:green}
\end{equation}

\noindent
This way, all the dependence on the  choice of vacuum is encoded into a
single harmonic function $h$. On the other hand,
as argued in section 3, the black holes created by the instanton
(\ref{eq:E-E}) are near-extremal Reissner-Nordstr\"{o}m in Rindler heat-baths.
That is, the Hawking radiation from the event horizon is in thermal
equilibrium with the heat bath associated with the
acceleration of the ambient region, showing that the proper vacuum to choose
is the Hartle-Hawking vacuum.

One important characteristic of the
Hartle-Hawking vacuum is that the potential
divergence of energy-momentum expectation values on both past and
future event horizon disappears. Moreover, we expect $h$ to be
independent of time coordinate, for the state is supposed to be in thermal
equilibrium. These two conditions are strong enough to fix $h$ almost
completely.
\begin{equation}
h=\pm F'(r_{+})\,z+C \label{eq:HM}
\end{equation}

\noindent
With this choice of the boundary condition, we find {\it semiclassical}
spacetimes  with regular event horizons of non-zero Hawking temperatures
(but with infinite ADM masses due to the heat bath), which is exactly what
one expects from the Hartle-Hawking vacuum \cite{Trivedi}\cite{Jaemo}.

To elaborate the procedure leading to the choice above, it is most convenient
to derive the one-loop energy-momentum tensor from a local form of the
Polyakov-Liouville action with the aid of an auxiliary scalar field $\psi$.

\begin{equation}
S_{PL}=S^{local}_{PL}\biggr\vert_{\nabla^2\psi =R^{(2)}}, \qquad
S^{local}_{LP}\equiv \frac{1}{96\pi}\int dx^2\sqrt{g^{(2)}}\,
\bigl( (\nabla \psi)^2+2\psi R^{(2)} \bigl)
\end{equation}
\begin{equation}
\Rightarrow \langle T_{ij} \rangle \sim - 2\nabla_{i}\nabla_{j}\psi +
2g_{ij}\nabla^2\psi
+\nabla_{i}\psi\nabla_{j}\psi-\frac{1}{2} g_{ij}(\nabla\psi)^2.
\end{equation}

\noindent
Solving the auxiliary equation with the help of the identity (\ref{eq:green}),
the energy-momentum can be rewritten in the light-cone coordinates
$x^{\pm}=t\pm z$, after a Wick rotation to the Minkowskian section
($d\tau^2=-dt^2$), as follows,
\begin{eqnarray}
\langle T_{\pm\pm} \rangle &\sim& -2\nabla_{\pm}\nabla_{\pm}(-\log F+h)
+\bigl(\nabla_{\pm}(-\log F+h) \bigr)^2 \nonumber \\
&=& -(\partial_{\pm}\log F)^2+2 \partial_{\pm}\partial_{\pm}\log F
    +(\partial_{\pm}h)^2-2\partial_{\pm}\partial_{\pm}h.
\end{eqnarray}

\noindent
For any finite temperature black hole, the  divergence (in a local
geodesic coordinate) on the event horizons is induced by the first term.
Choosing a $h=h(x^{-})$ to cancel this divergence on the future event horizon
leads to the familiar Hawking's radiation, corresponding to Unruh vacuum
\cite{Jaemo}\cite{CGHS}.
Choosing a $h=h(z)$ to cancel this divergence on both past and future event
horizons, therefore, must correspond to the
Hartle-Hawking vacuum. The resulting
choice $h=h(z)$ can be written most generally as in (\ref{eq:HM}).

This choice of vacuum is not only well-motivated physically but also
vital for the validity of the WKB approximation, in that the gravitational
backreaction to the quantum fluctuations is now well controlled.\footnote{
This has been emphasized previously in reference \cite{remnants}.}
 Otherwise, the
quantum fluctuations around the classical solution can no longer be regarded
as ``small'' and a systematic expansion based on the Euclidean instanton
would be ill-fated.

Note that there are some harmless ambiguities remaining. In particular,
an additive shift of the constant $C$ can be translated into an additive
shift of the topological term $a\rightarrow \tilde{a}$, and our ignorance
of $C$  will result in an unknown topological contribution $\sim
\chi_{\rm Euler}$ insensitive to the geometry and thus independent of $qB$.

Evaluating the Polyakov-Liouville action,
\begin{equation}
S_{PL}\rightarrow \frac{1}{96\pi} \int d\tau dz\,
(\partial_{z}^2 \log F)(\log F -h)\: +4\pi \tilde{a} \chi_{\rm Euler}.
\end{equation}

\noindent
Let us first evaluate with $h=F'(r_{+})z+\cdots$. Dropping the topological
terms $\sim \chi_{\rm Euler}$ which is finite as $qB\rightarrow 0$, we find
\begin{equation}
S_{PL}\biggr\vert_{on-shell} \rightarrow -\frac{\hbar}{96\pi T_{BH}}
\int_{r_{+}}^{r_B} dr\, \frac{F'(r)\,\bigl(F'(r)-F'(r_{+})\bigl)}{F(r)}.
\label{eq:main}
\end{equation}
If we had chosen $h=-F'(r_+)\,z+\cdots$ instead, the expression would have
changed only by another boundary contribution that is independent of the
geometry and again does not enter the leading $qB$ dependence.

The radial integration comes with the upper limit at $r_B$, for the
relevant geometry is again the truncated Euclidean black hole, but it
matters little owing to the rapidly vanishing behaviour of the integrand.
The resulting integral is finite for any $T_{BH}$, and is continuous
in the extremal limit.
\begin{equation}
\lim_{r_{+}\rightarrow r_{-}}\int^{\infty}_{r_{+}}\frac{F'(r)\,
\bigl(F'(r)-F'(r_{+})\bigl)}{F(r)} =\int^{\infty}_{q}\lim_{r_{+}\rightarrow
r_{-}}\frac{F'F'}{F}=\frac{4}{3q}
\end{equation}

\noindent
Again using the constraint $T_{BH}/\hbar =B/2\pi+\cdots$,
we arrive at the following one-loop corrected exponent,
\begin{equation}
-\frac{S_{E}}{\hbar}-W= -\frac{\pi q}{\hbar B }+\frac{N}{36qB}
+\cdots=-\frac{\pi q}{\hbar B}\biggl(1-\frac{N\hbar}{36\pi q^2}\biggr)+\cdots .
\label{eq:WKB1}
\end{equation}

\noindent
The ellipsis now denotes  terms from quantum fluctuations other than the
chargeless Callan-Rubakov modes as well as terms of higher order in $qB=qBG$.

Compared to the massive case (\ref{eq:Massive}), we find similar
behaviour, for there is no extra factor of $1/qB$ generated. Also
we can easily see that the analogue of $\sigma$ in
(\ref{eq:Massive}) is here given by a {\it negative} constant $-N/36\pi$,
corresponding to an enhancement of the rate for all small $B$. However,
within this large $N$ approximation ($N\hbar$ finite while $\hbar\rightarrow
0$) where one-loop matter contribution is dominant over that of internal
gravitons, we need to keep $q^2$ large in order to justify neglecting the
higher-loop gravitational contribution as well. Hence, the formula above
makes sense only for small $N\hbar/12\pi q^2$, and the one-loop correction
results in only a slight enhancement of the rate.

Before closing this section, we want to mention that the choice of boundary
condition actually does not affect the leading behaviour $W\simeq -N/36qB$.
Other choices result only in different boundary terms of order $(qB)^0$.
This is closely related to the fact that the functional form of the
Polyakov-Liouville action is completely determined by the conformal
anomaly, which is a local quantity.
Such a highly unusual character of chargeless Callan-Rubakov modes raises
a question whether they are ``generic'' enough to simulate actual
gravitational fluctuations that really contribute to the one-loop WKB.
In a sense, the derivation in this section should be regarded
more as an illustration than anything else.

\section{The Leading One-Loop Contributions in the Weak Field Limit:
Conformal 4-D Fluctuations}

Finding the four-dimensional effective action in its full non-local form is
a horrendous task, even for a simple quadratic matter action \cite{Covariant}.
On the other hand, what we actually need within the WKB approximation is just
values of the effective action on a specific family of metrics. In the
case at hand, the instanton has only two independent parameters $q$ and $B$,
or equivalently $q$ and $T_{BH}$.
Then, using the standard formula relating a variation of the effective
action to the energy-momentum expectation values, we should be able to reduce
the problem to that of finding $\langle T_{\alpha\beta} \rangle$ \cite{Stress}
in a given background.

\begin{equation}
\frac{\partial W}{\partial q}=\int dx^4\sqrt{g^{(4)}}\:\frac{\partial
g^{\alpha\beta}}{\partial q}\langle T_{\alpha\beta}\rangle, \label{eq:delW}
\end{equation}
where $W$ is the effective action evaluated on the two-parameter family
of the Euclidean metrics. The energy-momentum expectation values are to be
evaluated in the Hartle-Hawking vacuum, as was emphasized in section 3.

Now let us consider the approximate 4-D metric (\ref{eq:R-N}),
rewritten in the tortoise coordinate. It is of practical importance to
use the tortoise coordinate $z$ because the range of $z$ is independent of the
two parameters of the solution.
$$g^{(4)}=F(r(z))\,(d\tau^2+ dz^2)+
r(z)^2\,(d\theta^2+\sin^2\theta \,d\phi^2)+\cdots$$
Inserting this form of metric to the equation (\ref{eq:delW}) above,
\begin{equation}
-\partial_q W = \frac{4\pi \hbar}{T_{BH}}\int_{-\infty} dz\,r^2 F\,\biggl[
\frac{\partial_q F}{F}\,\langle T^{\tau}_{\tau}+T^{z}_{z} \rangle+2\frac{
\partial_q r}{r}\, \langle T^{\theta}_{\theta}+T^{\phi}_{\phi}\rangle\biggr]
+\cdots
\end{equation}
Inside the integral, the partial derivative $\partial_{q}$ should be carried
out with fixed $z$ and also fixed $B\simeq 2\pi T_{BH}/\hbar$.
To recover the one-loop contribution $W$, we just need to recall that the
only dimensionless combination of $q$ and  $B$ is $qB$, and that any
one-loop dimensionless physical quantity, without an explicit cut-off
dependence, must be a function of $qB$.

\vskip 5mm
Regrouping terms above, and using the spherical symmetry, we may simplify the
expression a little bit.
\begin{equation}
-\partial_q W =\frac{4\pi \hbar}{T_{BH}}\int_{r_+} dr \, r^2 \:\biggl[
               \frac{\partial_{q} F}{F}\biggr]\,
                \langle T^{\alpha}_{\alpha} \rangle \;
             +\frac{4\pi \hbar}{T_{BH}}\int_{r_+} dr\, r^2 \: \biggl[
                \frac{4\partial_q r}{r}-\frac{2\partial_q F}{F}\biggr]\,
                \langle T^{\theta}_{\theta} \rangle+\cdots. \label{eq:master}
\end{equation}
In general, both of the expectation values $\langle T_{\alpha}^{\alpha}
\rangle$ and  $\langle T_{\theta}^{\theta} \rangle$ depend on the
boundary condition strongly, and their precise behaviour is unknown,
especially in the extremal limit.

However, there are special cases where we can exploit a well-known field
theoretical fact concerning conformally coupled fluctuations.
While the trace of the energy-momentum is classically zero for such matter
fields, the one-loop expectation value thereof is nonvanishing due to the
conformal anomaly. The conformal anomaly $\langle T_{\alpha}^{\alpha}
\rangle $ is explicitly calculable and can be expressed as a polynomial
of the curvature tensor and its derivatives, and thus is completely local
\cite{Trace}.

To estimate the anomaly contribution to $W$, first write the coordinate
transformation between $r$ and the tortoise coordinate $z$, introducing
an arbitrary positive constant $a$.
\begin{equation}
z(r)=\int^r_{(a+1)r_+}\frac{d\tilde{r}}{F(\tilde{r})}\label{eq:defz}
\end{equation}
Defining the surface gravity of the horizon by $\kappa\equiv F'(r_+)=2\pi
T_{BH}/\hbar\simeq B$, then, $F$ is given by the following expansion in
$e^{2\kappa z}\simeq (r-r_+)/ar_+$  as $z\rightarrow -\infty$,
\begin{equation}
F(r(z))= 2\kappa a r_+\,e^{2\kappa z}+O(e^{4\kappa z}).
\end{equation}
Using the fact that $\partial_q \kappa\equiv 0\equiv \partial_q z$,
we then find the relative variation $\partial_q F/F$ approaches
$\partial_q r_+/r_+ \simeq 1/r_+ \simeq 1/q$ near the black hole horizon
provided that $\kappa$ is positive. In the asymptotic region $r\rightarrow
r_B \gg q$, on the other hand, it is not difficult to convince oneself
that $\partial_q F/F\simeq 1/r$ at most.

Then, since the value of the trace anomaly is of order $\sim 1/q^4$ near the
horizon and $\sim 1/r^5$ for large $r$, the first integral of
(\ref{eq:master}) is completely finite and contributes to $\partial_q W$ a
finite leading term of order $\sim \hbar/q^2T_{BH}\simeq 2\pi/q^2B$.
Hence, we conclude that the anomaly contribution to the one-loop
exponent is given by $\sim 1/qB$, just as in previous examples.
\vskip 5mm

However,  $\langle T_\theta^\theta \rangle$ strongly depends on the choice of
vacuum and shows a substantially different behaviour than the conformal
anomaly. For instance, in the Hartle-Hawking vacuum around a black hole,
it must approach a constant $\langle T_\theta^\theta \rangle_\infty \sim
T_{BH}^4\sim B^4$ at large $r$, for the quantum state there is that of
a heat bath at temperature $T_{BH}$.
Without a suitable infrared cut-off, this could result in
a divergence of $W$ as expressed above, requiring more careful analysis,
but as emphasized in section 3, the size of the bottom ``cup,''  $r_B
\sim (q B^{-2})^{1/3}$, effectively acts as an infrared cut-off, and such an
asymptotic heat bath of uniform $\langle T_\theta^\theta \rangle_\infty
\neq 0$ may contribute to the radial integral higher order
terms $< (qB)^1$ only.

Then, the only remaining part of (\ref{eq:master}) that can possibly
alter the leading weak field behaviour $W\sim 1/qB$ from the $\langle
T^\theta_\theta \rangle$ term near the horizon $r=r_+$. For this
remaining contribution, all we need to do is  determine the behaviour of
the following quantity $X(q,\kappa)$ as the extremal limit is approached
($\kappa \rightarrow 0$).
\begin{equation}
X \equiv \int_{r_+}^{2r_+} dr\, r^2 \: \biggl[
\frac{4\partial_q r}{r}-\frac{2\partial_q F}{F}\biggr]\,
\langle T^{\theta}_{\theta} \rangle, \label{eq:last}
\end{equation}
The potential divergence of $X$ can only originate at the lower limit of
this integral, and it is most important to understand the behaviour of
the integrand near the horizon for small $\kappa\simeq B >0$.

First let us consider the quantity inside the square brackets. Approaching
the lower bound, $\partial_q F/F$ converges to $\partial_q r_+/r_+$ provided
$\kappa >0$, as shown above, while $\partial_q r/r$ obviously approaches
$\partial_q r_+/r_+$ for any $\kappa \ge0$. Henceforth, the summed
fractional variation inside the square brackets approaches the finite value
$2\partial_q r_+/r_+\simeq 2/r_+\simeq 2/q$.
\begin{equation}
\lim_{r\rightarrow r_+}\;\biggl[\frac{4\partial_q r}{r}-\frac{2\partial_q
F}{F}\biggr] \simeq \frac{2}{q}\qquad \hbox{if\ }\kappa >0
\end{equation}
{}From this, we conclude that only rather strong divergence of $\langle
T^{\theta}_{\theta} \rangle \ge (r-r_+)^{-1}$ can cause singular behaviour
of $X$ as $\kappa \rightarrow 0$.

There exists extensive literature that deals with one-loop energy-momentum
tensor outside black holes.\footnote{See  \cite{Stress},  \cite{Trace},
\cite{Analytic}, \cite{Numerical}, and the references therein.}
Unfortunately, for the cases of
Reissner-Nordstr\"{o}m black holes, known analytic approximation schemes
\cite{Analytic}
are unreliable near the event horizon for any nonzero charge of the black
hole. But a numerical estimate using a mode sum was carried out by Anderson
and collaborators \cite{Numerical},
which seems to indicate that $\langle T_\theta^\theta \rangle$ remains
finite at the event horizon for both nonextremal and extremal cases. In fact,
they observed that the horizon value of $\langle T_\theta^\theta \rangle$
achieves a maximum at about $r_-/r_+ \simeq 0.92$ and decreases
again as the extremality $r_-/r_+ =1$ is approached.

If we take this numerical evidence seriously, we may immediately conclude
that $\lim_{\kappa\rightarrow 0}X$ is finite and  $\partial_q W$
behaves $\sim 1/q^2B+\cdots$ in the weak field limit $qB\rightarrow 0$.
Then, as in previous examples, the one-loop correction of the exponent
diverges as $1/qB$ and the net result is again that {\it the Schwinger
term is multiplicatively corrected by a fraction $\sim \hbar/q^2$.}

\vskip 5mm
Recently Trivedi demonstrated in the same S-wave approximation as in section
5 above that the local energy density, or equivalently $\langle T_{UU}
\rangle$ in the null Kruskal coordinates $(U,V)$, is divergent near an
extremal Reissner-Nordstr\"{o}m horizon \cite{Trivedi}. Yet this divergence
clearly does not enter the estimate (\ref{eq:main}), for the latter is
more or less determined by $\langle T_{UV} \rangle$ the conformal anomaly.
On the other hand, such a divergence in four dimensions could have entered
the four-dimensional estimate here, since (\ref{eq:master}) depends on more
than just the traced energy-momentum expectation. The natural question is
then how the potential divergence of $\langle T_{UU} \rangle$ affects
the behaviour of $\langle T_\theta^\theta \rangle$ and ultimately that
of $\partial_q W$. Is the finite behaviour of $X(q,\kappa)$ as $\kappa
\rightarrow 0$ consistent with such a divergence?

In appendix, we present a partial answer by exploiting the conservation
of energy-momentum to show that the divergence, if any, of angular components
is typically much weaker than that of $\langle T_{UU}\rangle$. Through
this analytical study, we shall demonstrate that Trivedi's divergence
is perfectly consistent with finite $\langle T_\theta^\theta \rangle$ and
that the quantity $X$ above is likely to be finite in the extremal limit.
Naturally, one wishes that more explicit and transparent calculations of
the energy-momentum were available. But even without such results, we
believe that the arguments given above combined with those in the appendix
provide strong evidences for the conclusion above.

\section{Conclusion}

In summary, we studied the weak field behaviour of the one-loop tunneling
rate of near-extremal magnetic black holes pair-production induced by
background magnetic fields. In particular, we considered the case of
Reissner-Nordstr\"{o}m black hole pair-production, mediated
by the Euclidean Ernst metric as the instanton.

Roughly, the instanton consists of two components: a single
Euclidean black hole without the asymptotic region beyond $r_B\sim
(qB^{-2})^{1/3}$ and a Euclidean Melvin space, glued together along the
transitional ``mouth'' region located at $\zeta\simeq B^{-1}$ and
$\rho \ll B^{-1}$, in terms of the Rindler-like coordinates (\ref{eq:Rindler})
of the Melvin space. Furthermore,
the Euclidean black hole is a near-extremal Reissner-Nordstr\"{o}m black hole
with Hawking temperature $T_{BH}\simeq \hbar B/2\pi$. The natural
boundary condition turned out to produce a variant of the
Hartle-Hawking vacuum that, outside the black hole region, behaves like
an ordinary Minkowski vacuum as seen by Rindler observers.

To simulate the one-loop contributions from quantized gravitational
fluctuations, we considered general chargeless quantum fluctuations governed
by various {\it elliptic} kinetic operators. In all cases considered,
we found the following weak field behaviour of the additive one-loop
correction $-W$ to the leading exponent $-S_E /\hbar$;
\begin{equation}
W\:\sim  \frac{1}{qB}\qquad\Longrightarrow
\qquad -\frac{S_E}{\hbar}-W =-\frac{\pi q}{\hbar B}\,\biggl(
1+\tilde{\sigma}\frac{\hbar}{q^2}\biggr)+\cdots, \label{eq:ATLAST}
\end{equation}
where $\tilde{\sigma}$ is a proportionality constant.\footnote{
What about the recently proposed ``geometric entropy''
\cite{GEOMETRIC} which happens to be proportional to $q^2$ and
also quadratically divergent? Where would such a divergence appear?
As Susskind and Uglum argued eloquently \cite{SUSS}, all the
quadratic divergence in the effective action multiplies the curvature
scalar (and a possible local boundary term involving the extrinsic
curvature)  only, and the quadratic divergence of ``geometric entropy'' can
always be traced back to the divergent renormalization of the inverse
gravitational constant $G^{-1}$. This can be easily seen by writing down
the effective action in the heat-kernel approach as in section 4.
Therefore, once we start with some ultraviolet finite theory of
gravity such as superstring theories, and as long as we express all
the quantities in terms of the renormalized couplings, no such
ultraviolet divergence may occur in the calculation of the tunneling
rate. As explicitly stated in section 4 and implicitly assumed in later
sections, we discarded all ultraviolet divergences by expressing
everything in terms of the renormalized couplings and in particular of the
renormalized $G^{-1}$ which is subsequently set to $1$.}

This behaviour is found for both massive and massless fluctuations after
appropriate renormalization procedures.  Also, we explicitly calculated
the additive contribution to $\tilde{\sigma}$ from $N$ chargeless
Callan-Rubakov modes, and found $-N/36\pi$.

\vskip 5mm
An immediate consequence of this is that there seems to be nothing
special about the weak field limit other than the fact that
$T_{BH}>0 $ becomes arbitrarily small.
In particular, the divergence of $W \sim 1/qB \sim \hbar/qT_{BH}$ in
the limit $T_{BH}\rightarrow 0$ results from the ever-increasing
periodicity $\hbar/T_{BH}$ of the Euclidean time coordinate
rather than from the diverging distance to the Euclidean black hole
horizon. Per unit Euclidean time, the contribution of quantum fluctuations
at one-loop level remains finite in this limit, in spite of the
ever-increasing 3-volume of the bottom ``cup'' which is in effect a truncated
near-extremal Euclidean black hole.

In fact, we believe that this leading one-loop correction can be explained
as a direct consequence of the semiclassically shifted mass-to-charge ratio
of near-extremal Reissner-Nordstr\"{o}m black holes \cite{Jaemo}.
To see this, it is necessary to restore the
{\it classical} mass $m\simeq q$ of near-extremal black holes in the
expressions (\ref{eq:Schwinger}) and (\ref{eq:ATLAST}),
\begin{eqnarray}
-\frac{S_E}{\hbar}&=&-\frac{\pi m^2}{\hbar qB}+\cdots, \\
-\frac{S_E}{\hbar}-W&=& -\frac{\pi m^2}{\hbar qB}\biggl(1+\tilde{\sigma}
\frac{\hbar}{q^2}\biggr)+\cdots . \nonumber
\end{eqnarray}
Note that the one-loop corrected expression would look like the original
Schwinger term, if we introduce a new parameter $\tilde{m}$ as follows:
\begin{equation}
-\frac{S_E}{\hbar}-W=-\frac{\pi \tilde{m}^2}{\hbar q B}+\cdots \quad
\hbox{with}\;\; \tilde{m}\simeq q\,(1+\frac{\tilde{\sigma}}{2}\,
\frac{\hbar}{q^2}). \label{eq:remass}
\end{equation}
This leads us to speculate that this particular one-loop correction
may represent a semiclassical effect that shifts near-extremal black hole
masses.

In order to verify this conjecture, it is necessary to ascertain such a
shift $m/q \rightarrow \tilde{m}/q$ in some other way. In particular,
given a fixed magnetic charge $q$, the lower bound of the
Reissner-Nordstr\"{o}m black hole masses, or equivalently the mass of the
extremal black hole, must be shown to be modified accordingly.
Exactly how does such a semiclassical correction arise?

Although the extremal black hole is known to have vanishing Hawking
temperature and does not emit the usual late-time thermal radiation, this
tells us nothing about transient behaviours before the state settles
down to a steady state. In fact, if we imagine an idealized
gravitational collapse that produces an extremal black hole \cite{Jaemo},
it is easy to see that there could be some transient quantum radiations
of finite integrated flux. The subsequent correction to the
mass-to-charge ratio was first demonstrated and estimated in reference
\cite{Jaemo}, where $N$ chargeless Callan-Rubakov modes are quantized around
Reissner-Nordstr\"{o}m black holes.

Both analytic and numerical studies,
again utilizing the power of Polyakov-Liouville effective action,
revealed that the semiclassically corrected extremal black holes
obey the following mass-to-charge ratio, when $N\hbar/12\pi q^2$ is small;
\begin{equation}
\frac{\rm mass}{\rm charge}\:\simeq 1-\frac{N\hbar}{72\pi q^2} .
\end{equation}
Accordingly, the lower bound of black hole masses is reduced in the same
fashion, and in particular near-extremal black holes now have the mass
approximately of $q-N\hbar/72\pi q$ rather than of $q$.
This is certainly consistent with the above interpretation (\ref{eq:remass}),
since the contribution to $\tilde{\sigma}/2$ from  $N$ chargeless
Callan-Rubakov modes is indeed $-N/72\pi$ as demonstrated in section 5.
Therefore, the one-loop correction we found merely reflects the fact
the mass of the charged object is modified at the one-loop level.
The analogous shifts of the mass-to-charge ratio due to general
quantum fluctuations are yet to be estimated, but it is reasonable
to expect the same interpretation to hold for other cases as well.

Seen the other way around, this observation also provides us with a new and
more systematic way of calculating the semiclassical mass-to-charge ratio
in the presence of fully four-dimensional quantum fluctuations. In reference
\cite{Jaemo}, the ratio was obtained by evolving an initially empty
energy-momentum expectation in accordance with the energy conservation,
but the method employed there is not easily generalized beyond the
particular case of chargeless Callan-Rubakov modes. On the other hand,
the formulae (\ref{eq:ATLAST}) and (\ref{eq:remass}) enable
us to convert this complicated
evolution problem to a static one, for all we need is a reasonably accurate
estimate of the energy-momentum expectations in the {\it Hartle-Hawking} vacuum
around near-extremal black holes.

\vskip 5mm
Here we discussed only the cases where the pair-produced objects are
{\it near-extremal} Reissner-Nordstr\"{o}m black holes, while
there are other known species of instantons \cite{GAUNT}\cite{GG}.
In particular, an especially interesting variant exists
that pair-produces strictly extremal Reissner-Nordstr\"{o}m black holes
rather than near-extremal ones. Moreover, this new class of instantons
are qualitatively different, in that the matching condition of
type (\ref{eq:R-H}) is absent: The pair-production of strictly extremal
black holes, if any, seems genuinely different from that of
the nonextremal ones, however close to the actual extremality the
latter might be. A central issue in extending the current analysis
to the strictly extremal cases would be how the natural vacuum, analogous
to the one we found and utilized here for {\it wormhole} type instantons,
behaves near the extremal horizon. The mismatch of the two surface
gravities seems to suggest a semiclassical instability that involves
strongly divergent energy-momentum near the black hole horizon and a large
gravitational backreaction thereof. It would be interesting to see if
the {\it extremal} type instantons allow semiclassical estimates that
are reliable at one-loop and beyond.

\vskip 5mm
In any case, the main result (\ref{eq:remass})
is most sensible in that it is exactly what
one would expect for pair-production of ordinary charge particles.
But at the same time it is rather disappointing. Despite the
ever-increasing size of the bottom "cup" that is uniquely associated
with the near-extremal black hole pair-production, even the strongest
correction in the weak field limit can be explained away and does
not lead to new interesting physics. But are we completely sure that
we did not leave out some essential physics in the course of carrying
out various approximations and simplifications?

It is always a possibility that some nontrivial and large physical
effects are hidden in
higher order terms, especially in $B$ independent one-loop contributions
that must include a correction to the Bekenstein-Hawking entropy.
Unlike our calculations here, unfortunately, the estimate of this
next-to-leading order in $qB$ appears a lot more sensitive to how
one treats the transitional "mouth" region, and thus is much more
difficult to carry out. For example, we have been rather cavalier
about the boundary at $r=r_B$ of the {\it truncated} black hole
and pretended that it is infinitely far away, which is justifiable
only as far as the leading $qB$ behaviour found above is concerned.

Of course, the most problematic aspect of the Euclidean approach here
is  the fact that Euclidean path integral is ill-defined with the
Einstein-Hilbert action, which we circumvented by substituting general
{\it matter} fields for actual gravitational fluctuations. In other words,
the genuine operators governing small gravitational fluctuations are {\it
not} even elliptic, and in particular possess infinite number of
zero-eigenmodes, not to mention infinite number of negative eigenmodes.
Note that by substituting in general {\it matter} fluctuations as above,
we in effect concentrated on positive eigenmodes-modes only,
Although this is not necessarily in conflict with our desire to find the
leading infrared behaviour in the weak field limit, we surely
neglected the possible degeneracy, if any, of black hole quantum
states. Therefore it is a matter of some urgency to understand how to
perform a next-to-leading WKB approximation in the presence of
gravitational degrees of freedom.

\vskip 5mm
Before closing, we want to bring up another related problem, somewhat
tangential to our purpose here but nevertheless rather intriguing on its own.
In this article, we needed to consider the vacuum in the Euclidean sector
only, which simplified the discussion quite a bit. An interesting question
to ask is what does the natural vacuum look like after the pair-production,
that is, along the Minkowskian sector where two oppositely charged black
holes are uniformly accelerating away from each other with $A \simeq B$.
The discussion in section 3 suggests that, as far as the co-accelerating
Rindler observers are concerned, the same thermal equilibrium
between Hawking radiation and Rindler heat bath must be maintained and
these pair-produced black holes are {\it semiclassically} stable.
But what does an inertial observer see? Does he find Doppler-shifted
Hawking radiations from each nonextremal black hole? Or, since the black
holes are in equilibrium with Rindler heat baths and do
not seem to lose any energy, should he agree that no quantum
radiation emerges from the black holes?
(This problem of accelerating black hole in the Rindler heat bath was
recognized also by Dowker and collaborators in reference \cite{GG}.)

Although the situation looks superficially similar to the case of
accelerating charge, where inertial observers inevitably find the
Bremsstrahlung \cite{BOULWARE}\cite{HIGUCHI}, the analogy breaks down
due to the entirely different nature of the respective radiations.
For instance, unlike the case of accelerating charge \cite{RANDERS},
we cannot expect the interference between the background fields
and the radiation to explain the energy conservation at all time,
since the Hawking radiation is completely universal in its composition
while there exist only gravitational and electromagnetic background fields.

While we are tempted by the simplest conclusion that these black holes
do not emit any Hawking radiation unless the fine-tuned uniform acceleration
is disrupted, it is by no means clear that some subtlety in measuring the
mass of such black holes may not explain the two potentially conflicting
reports by inerial and co-accelerating observers. It should be a most
interesting exercise to obtain the complete semiclassical picture of this
system, which probably requires deeper understanding of both classical
and semiclassical physics of black holes.

\vskip 1cm

\centerline{\bf ACKNOWLEDGMENT }

\vskip 5mm

I wish to thank Dr. S. Trivedi and Prof. J. Preskill for several invaluable
conversations as well as for suggesting this problem in the first place.
This work is completed during a visit at the Center for Theoretical
Physics of Seoul National University, where I enjoyed the hospitality
of its members and in particular of Prof. Choonkyu Lee. This work was
supported in part by D.O.E. Grant No. DE-FG03-92-ER40701.

\vskip 1cm

\leftline{\Large\bf Appendix: Taming $\langle T_\theta^\theta
\rangle $ at the Event Horizon.}
\vskip 5mm\noindent
In this appendix, we want to study $\langle T^\theta_\theta\rangle$ near
the black hole horizon. In the Hartle-Hawking vacuum around a nonextremal
black hole, $\langle T^\theta_\theta \rangle$ is expected to remain finite
approaching the event horizon. But it is rather uncertain what happens in
the extremal limit, in particular in view of the two-dimensional results
by Trivedi that $\langle T_{UU}\rangle$ diverges at the extremal horizon
\cite{Trivedi}. The objective of this appendix is to place an upper bound
on the possible divergence of $\langle T^\theta_\theta \rangle$ for the
purpose of estimating the quantity $X$ of (\ref{eq:last}). The results here
are meant to be complementary to the numerical estimates of reference
\cite{Numerical}.

First, recall that the energy-momentum must satisfy the
conservation equations, being Noether currents associated with the
diffeomorphism invariance.
\begin{equation}
\nabla_{\beta} T_{\alpha}^{\beta}\equiv 0
\end{equation}
Given any static and spherically symmetric expectation values $\langle
T_{\alpha\beta} \rangle$, such as expected in the Hartle-Hawking vacuum, these
constraints can be reduced to a first order differential equation for
$\langle T_{z}^{z} \rangle$.\footnote{We are indebted to Jaemo Park
for pointing  this out.}

Solving it, one finds the following relationship between different components
of the energy-momentum expectation.
\begin{eqnarray}
\langle -T^{t}_{t}+T^{z}_{z} \rangle=\langle -T^{\tau}_{\tau}+T^{z}_{z}\rangle
&=&\biggl[\: -\,\langle T_{\alpha}^{\alpha}\rangle+\frac{1}{r^2F}
\int^r_{r_0}  dr\:
r^2\frac{dF}{dr} \,\langle T^{\alpha}_{\alpha} \rangle\: \biggr] \nonumber \\
&+&\biggl[\: 2\,\langle T_{\theta}^{\theta} \rangle -\frac{2}{r^2F}
\int^r_{r_0} dr\:
(r^2\frac{dF}{dr}-2rF) \,\langle T_{\theta}^{\theta} \rangle\: \biggr]
\label{eq:master2}
\end{eqnarray}

\noindent
Obviously, the lower limit $r_0$ for each integral must be chosen in
accordance with the expected behaviour of the vacuum.

For instance, in the
Boulware vacuum around a nonextremal black hole, $\langle T^\alpha_\beta
\rangle $ must vanish very rapidly as $r\rightarrow \infty$, and the only
choice consistent with this is $r_0=\infty$.
On the other hand, the left-hand-side is related to the energy-density as seen
by inertia observers crossing the future event horizon. Calling the in-going
Kruskal coordinate $U$, this energy-density at the horizon scales like
\begin{equation}
\langle T_{UU}\rangle \sim \frac{\langle -T_{t}^{t}+T_{z}^{z} \rangle}{F},
\end{equation}
the leading divergence of which at $F=0$ disappears only
if the lower limit of the integrals in (\ref{eq:master2}) is at the event
horizon. Therefore, in contrast with the Boulware vacuum, the Hartle-Hawking
vacuum requires $r_0=r_+$.

In any case, the upshot is that now we can relate the angular components of
one-loop energy-momentum at the event horizon to the leading behaviour of
the energy-density.

\vskip 5mm
Using this constraint, it is easy to set an upper bound on the potential
divergence of  $\langle T^\theta_\theta \rangle$ in the {\it Boulware}
vacuum by exploiting this constraint, as follows.
Assuming that $\langle T_\theta^\theta \rangle
\sim (r-r_+)^{-m}$ near the horizon, we find from (\ref{eq:master2})
with $r_0=\infty$ the following leading divergence of the energy-density.
\begin{equation}
\langle T_{UU} \rangle_{\rm nonextremal}^{\rm Boulware}\sim \left\{
\begin{array}{ll} (r-r_+)^{-2} \sim U^{-2}&\mbox{if $m\le 1$} \\
                  (r-r_+)^{-1-m} \sim U^{-1-m}
&\mbox{otherwise} \end{array}\right.  \label{eq:boulware}
\end{equation}
Here we have used the facts that $\langle T_\beta^\alpha \rangle^{\rm Boulware}
$ vanishes very rapidly in the asymptotic region and that $F\sim (r-r_+)$
near the nonextremal event horizon.
But, the divergence of $\langle T_{UU}\rangle^{\rm
Boulware}_{\rm nonextremal}$ is closely related to Hawking's late time thermal
radiation. For a realistic black hole, the future event horizon is smooth as
seen by freely infalling observers, and the corresponding {\it Unruh} vacuum
must have an extra contribution to $\langle T_{UU} \rangle$ that cancels
the divergence that appears in the Boulware vacuum. This extra
contribution is predominantly a function of $U$, and thus propagates outward
to future null-infinity. The resulting energy flux can be estimated from
(\ref{eq:boulware}) combined with appropriate coordinate transformation,
\begin{equation} \Longrightarrow\qquad
\langle T_{uu} \rangle_{\rm nonextremal}^{\rm Unruh}\biggr\vert_{u,r
\rightarrow \infty} \sim \left\{
\begin{array}{ll}  1/r^2&\mbox{if $m\le 1$} \\
                  e^{(m-1)\kappa u}/r^2&\mbox{otherwise.} \end{array}\right.
\end{equation}
Here $u\equiv t-z$ is the asymptotic retarded time, and $\kappa$ is the
surface gravity of the event horizon. The exponentially divergent behaviour
of the second ($m> 1$) is clearly unacceptable, and in fact, we already
know that the radiation flux from the nonextremal black hole is thermal
and must be independent of the retarded time $u$ \cite{Hawking}.
This requires $m$ not to be
larger than $1$, putting an upper bound on the possible divergence of
of $\langle T_\theta^\theta \rangle$ in the Boulware vacuum.
\begin{equation}
\langle T_\theta^\theta \rangle^{\rm Boulware}_{\rm nonextremal} \le
(r-r_+)^{-1}\sim U^{-1} \qquad\Longleftarrow \qquad
\langle T_{UU} \rangle^{\rm Boulware}_{\rm nonextremal} \sim (r-r_+)^{-2}
\sim U^{-2}
\end{equation}
Note that the conservation of energy-momentum restricts the possible
divergence of the angular components to be much weaker that that of
$\langle T_{UU}\rangle$.

\vskip 5mm
Similar analysis can be carried out for the {\it  Hartle-Hawking} vacuum, too.
In this case, $\langle T_{\alpha}^{\beta}\rangle$, in a local geodesic
coordinate, are expected to be
finite at the nonextremal horizon, and this is certainly consistent with
the constraint above. To see this, we perform integrations by part to obtain
the following relationship in the Hartle-Hawking vacuum, which is valid
provided that both  $F\, \langle T_\alpha^\alpha \rangle$ and $F\,\langle
T_\theta^\theta \rangle$ vanish at $r=r_+$,
\begin{equation}
\langle T_{UU}\rangle \sim \frac{1}{r^2 F^2}
\int^r_{r_+} dr\; F\, \biggl\{ \frac{d}{dr}[2r^2\langle T_\theta^\theta
\rangle -r^2 \langle T^\alpha_\alpha \rangle] +4r\langle T_\theta^\theta
\rangle \biggr\}. \label{eq:mild}
\end{equation}
For regular $\langle T^\theta_\theta \rangle$, the quantity inside the
curly-bracket is finite at $r=r_+$, and the definite integral vanishes
quadratically $\sim (r-r_+)^2$ near the nonextremal horizon. Taking into
account the fact that $F$ vanishes linearly $\sim (r-r_+)$, we conclude
\begin{equation}
\langle T_{\theta}^\theta \rangle_{\rm nonextremal}^{\rm Hartle-Hawking}\sim 1
\qquad \Longleftrightarrow \qquad
\langle T_{UU} \rangle_{\rm nonextremal}^{\rm Hartle-Hawking}\sim 1.
\end{equation}
This gives us more confidence that $X(q,\kappa >0)$ of (\ref{eq:last})
is really finite.
Now what about the extremal limit $\kappa \rightarrow 0$? As mentioned in
section 6, the numerical results of Anderson et al., indicate that $\langle
T_\theta^\theta \rangle$ at horizon actually decreases as $\kappa\rightarrow 0$
in near-extremal cases ($r_-/r_+>0.92$), which implies finite limit of
$X(q,\kappa\rightarrow 0)$. But, to be extra cautious, let us try to
determine the behaviour of $X$ in the strictly extremal case $\kappa=0$.

\vskip 5mm
Now we want to extend the above analysis to the extremal limit of the
Hartle-Hawking vacuum.\footnote{Since the Boulware and the Hartle-Hawking
vacua coincide in the extremal limit, we will drop the superscript
``Hartle-Hawking'' in the following
formulae.} Suppose that $\langle T_\theta^\theta
\rangle$ approaches a finite nonzero value near the extremal horizon.
Recalling that $F$ now vanishes quadratically $\sim (r-r_+)^2$, we find from
(\ref{eq:mild}) the following divergence of the energy-density.
\begin{equation}
\langle T_{UU} \rangle_{\rm extremal}\sim (r-r_+)^{-1} \sim U^{-1}.
\label{eq:TR_EX}
\end{equation}
What if $\langle T_\theta^\theta
\rangle$ is logarithmically divergent $\sim \log(r-r_+)\,$? In this case,
the differentiation of the square bracket in (\ref{eq:mild}) produces an extra
$1/(r-r_+)$ term that generates the leading term of the integral $\sim
(r-r_+)^2$. As a result, we find
\begin{equation}
\langle T_{UU} \rangle_{\rm extremal}\sim (r-r_+)^{-2} \sim U^{-2}.
\end{equation}
Finally, the case of $\langle T_\theta^\theta \rangle \sim (r-r_+)^{-n}$
for general $n$ can be studied similarly from the original equation
(\ref{eq:master2}) with $r_0=r_+$, which yields,
\begin{equation}
\langle T_{UU} \rangle_{\rm extremal}\sim (r-r_+)^{-2-n} \sim U^{-2-n}.
\end{equation}
The proportionality constant here vanishes if and only if $n=0$. Again
we can see that, owing to the conservation of energy-momentum, the
divergence of $\langle T_{UU}\rangle$ is much more severe than that of the
angular components. In particular, the divergence shown in (\ref{eq:TR_EX})
is similar to what Trivedi found in two-dimensional context, but the
corresponding $\langle T_\theta^\theta \rangle$ is perfectly finite.

It is rather difficult to imagine why $\langle T_{UU} \rangle_{\rm extremal}
=\lim \langle T_{UU} \rangle_{\rm nonextremal}^{\rm Hartle-Hawking}$ should
be more divergent than $\langle T_{UU} \rangle^{\rm Boulware}_{\rm nonextremal
}$, but, even allowing such strong divergence as $\langle T_{UU}\rangle_{\rm
extremal}\sim U^{-4}$, we find the quantity $X(q,\kappa=0)$ completely finite.

This can be seen from the fact that, in the strict extremal limit ($\kappa=0$),
(\ref{eq:defz}) with extremal $F$ implies that $\partial_q F/F$ converges
to $2\partial_q r_+/r_+$ instead of $\partial_q r_+/r_+$, which translates into
\begin{equation}
\biggl[\frac{4\partial_q r}{r}-\frac{2\partial_q F}{F}\biggr] \sim
\frac{(r-r_+)^2}{r_+^2}\qquad \hbox{if\ }\kappa =0.
\end{equation}
The definition in (\ref{eq:last}), then, shows that $X(q,0)$ can be infinite
only if the divergence of $\langle T_\theta^\theta  \rangle$
is at least as strong as
$\sim (r-r_+)^{-3}$. From the general relationships shown above, it is clear
that $X(q,0)$ remains finite as long as $\langle T_{UU}\rangle$
is less divergent than $\sim U^{-5}$.

\vskip 5mm
In this appendix,  relying on purely analytical methods, we provided some
additional evidence that the pivotal quantity $X$ remains finite in the
extremal limit.


\begin{thebibliography}{99}
\bibitem{ERNST}{D. Garfinkle and A. Strominger, Phys. Lett. 256B (1991) 146.}
\bibitem{GGS}{D. Garfinkle, S. Giddings and A. Strominger, Phys. Rev. D49
(1994) 958.}
\bibitem{GAUNT}{F. Dowker, J. P. Gauntlett, D. A. Kastor and J. Traschen,
Phys. Rev. D49 (1994) 2909.}
\bibitem{GG}{F. Dowker, J. P. Gauntlett, S. B. Giddings and G. T. Horowitz,
{\it On Pair Creation of Extremal Black Holes and Kaluza-Klein Monopoles},
EFI-93-74, UCSBTH-93-38, hep-th/9312172.}
\bibitem{Info}{S. W. Hawking, Phys. Rev. D14 (1976) 2460; Comm. Math. Phys.
87 (1982) 395.}
\bibitem{remnants}{See for example S. Giddings, Phys. Rev. D49 (1994) 947;
{\it Comments on Information Loss and Remnants}, UCSBTH-93-35, hep-th/9310101.}
\bibitem{BEK}{J. D. Bekenstein, Phys. Rev. D7 (1973) 2333; {\it ibid.} D9
(1974) 3292; G. W. Gibbons and S. W. Hawking, {\it ibid.} D15 (1977) 2752.}
\bibitem{ColeWKB}{S. Coleman, {\it Aspects of Symmetry}, Cambridge
University Press, 1985.}
\bibitem{MONO}{I. K. Affleck and N. S. Manton, Nucl. Phys. B194 (1982) 38.}
\bibitem{Trivedi2}{A. Strominger and S. Trivedi, Phys. Rev. D48 (1993) 5778.}
\bibitem{SD}{See G. A. Vilkovisky, Phys. Rep. 119 (1985) 1.}
\bibitem{Callan}{C. G. Callan, Phys. Rev. D26 (1982) 2058; {\it ibid.} D25
(1982) 2141; V. A. Rubakov, Nucl. Phys. B203 (1982) 311.}
\bibitem{Alford}{M. G. Alford and  A. Strominger, Phys. Rev. Lett. 69
(1992) 563.}
\bibitem{Holzhey}{C. F. E Holzhey and F. Wilczek, Nucl. Phys. B380 (1992) 447.}
\bibitem{theta}{P. Yi, Phys. Rev. D49 (1994)  5295.}
\bibitem{Jaemo}{J. Park and P. Yi, Phys. Lett. 317B (1993) 41.}
\bibitem{Unruh}{S. A. Fulling, Phys. Rev. D7 (1973) 2850;
P. C. W. Davies, J. Phys. A8 (1975) 609; W. G. Unruh, Phys. Rev. D14 (1976)
870.}
\bibitem{Wald}{W. G. Unruh and R. M. Wald, Phys. Rev. D29 (1984) 1047.}
\bibitem{Hawking}{S. W. Hawking, Comm. Math. Phys. 43 (1975) 199.}
\bibitem{Heat}{An elementary introduction to the heat kernel can be found
in J. Roe, {\it Elliptic Operators, Topology and Asymptotic
Methods}, Longman Scientific \& Technical, 1988.}
\bibitem{Covariant}{A. O. Barvinsky and G. A. Vilkovisky, Nucl. Phys. B282
(1987) 163; {\it ibid.} B333 (1990) 471; references therein.}
\bibitem{Polyakov}{A. M. Polyakov, Phys. Lett. 103B (1981) 207.}
\bibitem{Trivedi}{S. Trivedi, Phys. Rev. D47 (1993) 4233.}
\bibitem{CGHS}{C. G. Callan, S. Giddings, J. Harvey and A. Strominger,
Phys. Rev. D45 (1992) R1005.}

\bibitem{Stress}{S. M. Christensen, Phys. Rev. D 14 (1976) 2490.}
\bibitem{Trace}{For a comprehensive review of quantum field theory in curved
spacetime, see N. D. Birrell and P. C. W. Davies, {\it Quantum Fields in
Curved Space}, Cambridge University Press, 1982}
\bibitem{Analytic}{V. P. Frolov and A. I. Zel'nikov, Phys. Rev D35 (1987)
3031.}
\bibitem{Numerical}{P. A. Anderson, W. A. Hiscock and D. A. Samuel,
Phys. Rev. Lett. 70 (1993) 1739; references therein.}
\bibitem{GEOMETRIC}{C. G. Callan and F. Wilczek, {\it On Geometric Entropy},
IASSNS-HEP-93-87, hep-th/9401072.}
\bibitem{SUSS}{L. Susskind and J. Uglum, {\it Black Hole Entropy in
Canonical Gravity and Superstring Theory}, SU-ITP-94-1, hep-th/9401070.}
\bibitem{BOULWARE}{D. G. Boulware, Ann. Phys. (New York) 124 (1980) 169.}
\bibitem{HIGUCHI}{A. Higuchi, G. E. A. Matsa, and D. Sudarsky, Phys. Rev.
D45 (1992) R3308; {\it ibid.} D46 (1992) 3450.}
\bibitem{RANDERS}{S. Coleman, unpublished.}





\end{thebibliography}
\end{document}